\documentclass[aps,prb,twocolumn,superscriptaddress,showpacs]{revtex4-1}
\usepackage{graphicx}
\usepackage{amsmath}
\usepackage{graphics}
\usepackage{exscale}
\usepackage{epsfig}

\begin{document}
\title[]{Fermi  Liquids and the Luttinger Integral}
\author{ O. J. Curtin}
\affiliation{Dept. of Mathematics, Imperial College, London SW7 2AZ, UK.}
\author{Y. Nishikawa}
\affiliation{Dept. of Physics, Osaka City University, Sumiyoshi-ku,
Osaka 558-8585 Japan.}
\author{ A. C. Hewson}
\affiliation{Dept. of Mathematics, Imperial College, London SW7 2AZ, UK.}
\author{D. J. G. Crow}
\affiliation{Dept. of Mathematics, Imperial College, London SW7 2AZ, UK.}

\date{\today}


\begin{abstract}
The Luttinger Theorem, which relates the electron density to the volume of the Fermi surface
in an itinerant electron system,  is taken to be one of the
essential features of a Fermi liquid. The microscopic derivation of this result  
depends on the vanishing of a certain integral, the Luttinger integral $I_{\rm L}$,
which is also the basis of the Friedel sum rule for impurity models,  relating 
the impurity occupation number to the scattering phase shift of the conduction electrons. 
It is known that non-zero values of  $I_{\rm L}$ with $I_{\rm L}=\pm\pi/2$,
 occur in impurity models in phases with non-analytic  low energy scattering, classified as singular
Fermi liquids. Here we show
the same values,  $I_{\rm L}=\pm\pi/2$, occur in an  impurity model in phases 
with  regular low energy Fermi liquid behavior. Consequently the Luttinger integral
can be taken to characterize these phases,  and the quantum critical
points separating them interpreted as topological.  
\end{abstract}
\maketitle
The characteristic feature of a Fermi liquid is that the low energy behavior can be understood in terms
of interacting quasiparticles and their collective excitations. In the Landau formulation
these are taken to be in 1-1 correspondence with those of the non-interacting system, such that the volume of 
the Fermi surface in the interacting system gives the electron density. Using the results of 
 Luttinger\cite{Lut60,Lut61} in his microscopic derivation of Fermi liquid theory we can define quasiparticles which have
an infinite lifetime.
 We consider a three dimensional
lattice system with Bloch states with energy $\epsilon({\bf k})$ and a single electron Green's function
$G({\bf k},\omega)$ with a self-energy at zero temperature $\Sigma({\bf k},\omega)$ due to interactions.
We rewrite the self-energy in the form\cite{Hew93,Hew16},
\begin{equation}\Sigma({\bf k},\omega)=\Sigma({\bf k}_{\rm F},0)+\omega\Sigma'({\bf k}_{\rm F},0)+\Sigma^{\rm rem}({\bf k},\omega),\end{equation}
where the Fermi wavevectors ${\bf k}_{\rm F}$, and hence the Fermi surface, are defined by the condition $\epsilon({\bf k}_{\rm F})+\Sigma({\bf k}_{\rm F},0)=0$
 and $\Sigma^{\rm rem}({\bf k},\omega)$ is the remainder term. From Luttinger's results\cite{Lut61} we take the $\omega$-derivative $\Sigma'({\bf k}_{\rm F},0)$ to be real and $\Sigma^{\rm rem}({\bf k}_{\rm F},\omega)\sim \omega^2$ as $\omega\to 0$, giving
\begin{equation}G({\bf k},\omega)=\frac{z({\bf k}_{\rm F})}{\omega-\tilde\epsilon({\bf k})-\tilde\Sigma({\bf k},\omega)},\end{equation}
where $\tilde\epsilon({\bf k})=z({\bf k}_{\rm F})(\epsilon({\bf k})-\epsilon({\bf k}_{\rm F}))$, $\tilde\Sigma({\bf k},\omega)=z({\bf k}_{\rm F})\Sigma^{\rm rem}({\bf k},\omega)$ and $z({\bf k}_{\rm F})=(1-\Sigma'({\bf k}_{\rm F},0))^{-1}$ . We can define a free
quasiparticle Green's function, $\tilde G_0({\bf k},\omega)$,
\begin{equation}\tilde G_0({\bf k},\omega)=\frac{1}{\omega-\tilde\epsilon({\bf k})}.\end{equation}
The Luttinger theorem is then equivalent to the statement that the total number of electrons corresponds to
an integration of the {\em free quasiparticle} spectral density over all the states ($\tilde\epsilon({\bf k})<0$) up to the Fermi level $\omega=0$, provided the integral
\begin{equation}I_{\rm L}= {\rm Im} \int_{-\infty}^{0}  \sum_{\bf k}\left(G({\bf k},\omega)\frac{\partial\Sigma({\bf k},\omega)}{\partial\omega}\right)d\omega
=0.\end{equation}\par
 
Essentially the same condition applies for the Friedel sum rule, which  gives the number of impurity electrons
$n_{\rm imp}$ in terms of the phase shift $\eta$ of the conduction electrons\cite{Fri56}.
 For example, for the Anderson impurity model with an
impurity d-level $\epsilon_d$ hybridized with conduction band electrons $\epsilon_k$, with a hybridization
matrix element $V_k$, this takes the form,
\begin{equation} n_{d}= \frac{2}{\pi}\eta+\frac{2}{ \pi}I_{\rm L},\label{lfsr}\end{equation}
where for an Anderson model with a flat wide conduction band $n_{\rm imp}=n_d$, with
\begin{equation} \eta= \frac{\pi}{2}-{\rm tan}^{-1}\left(\frac{\epsilon_d+\Sigma_R(0)}{\Delta}\right),\quad I_{\rm L}={\rm Im}\int G_d(\omega)\frac{\partial\Sigma(\omega)}{\partial\omega}d\omega,
\end{equation}
where  $\Delta=\pi\sum_k|V_{k}|^2\delta(\epsilon_{k})$, and  $G_{d}(\omega)$ is the impurity d-Green's function,  $(G_{d}(\omega))^{-1}=(\omega+i\Delta\,{\rm sgn}(\omega)-\epsilon_{d}-\Sigma(\omega))$, where $\Sigma_R(\omega)$ is the real part of the self-energy $\Sigma(\omega)$. The Friedel sum rule corresponds to the case\cite{Lan66} where the occupation is determined entirely by the phase shift, i.e. $I_{\rm L}=0$.
\par 
The question has been raised over a number of years as to whether Luttinger's theorem holds in certain regimes   of
models used to describe  strongly correlated electron systems\cite{CMS95pre,LSGB95,PLS98,CZ99,KO03,KF05}.
There is also recent experimental evidence\cite{MKZWW12} in the underdoped phase of the cuprate superconductors that the volume of the Fermi surface corresponds  not to the total electron number $1-p$ but to the doping level $p$.
Without definitive results for models of these systems the question remains open.  There are, however,
exact results for many impurity models where this question can be put to the test. It has been found
that  there are some impurity systems\cite{LWG09}, such as an underscreened Kondo model\cite{MLK11,LTG14}, and for certain parameter
regimes in models of a triangular arrangement of quantum dots\cite{MJGL13,TBR16}, where Eqn. (\ref{lfsr}) is only satisfied if
$I_{\rm L}$ takes values $\pm\pi/2$. The low energy fixed point in a numerical renormalization group (NRG) for these systems corresponds to free fermions, with leading irrelevant terms that are non-analytic in $\omega$,
taking the form $1/({\rm ln}(\omega/T_{\rm K}))^2$, where $T_{\rm K}$ is a Kondo temperature. As a consequence these have been classified as {\em singular} Fermi liquids.\par
We show here the existence of  phases of an impurity model with the low energy behavior corresponding
to well defined quasiparticles together with interaction terms that give the usual  low energy frequency and temperature Fermi liquid scattering effects
of order $\omega^2$ and $T^2$  but with {\em non-zero values of the Luttinger integral}, $I_{\rm L}=\pm\pi/2$.\par 
The model we consider describes two quantum dots or impurities coupled
by an antiferromagnetic exchange and direct term, with a Hamilonian  
${\cal H}=\sum_{\alpha=1,2}{\cal H}_\alpha+{\cal H}_{12}$,
 with ${\cal H}_\alpha$ corresponding to an individual  Anderson impurity model
with channel index $\alpha$,
\begin{eqnarray}
&&{\cal H}_\alpha=\sum_{\sigma}\epsilon_{d,\alpha}d^{\dagger}_{\alpha,\sigma}d^{}_{\alpha,\sigma}+\sum_{k,\sigma}\epsilon_{k,\alpha}
c^{\dagger}_{k, \alpha,\sigma} c^{}_{k, \alpha,\sigma}\label{model1a} \\
&&+\sum_{k,\sigma} (V_{k,\alpha} d^{\dagger}_{\alpha,\sigma} c^{}_{k,\alpha,\sigma}
+ {\rm h.c.})+ U_\alpha n_{d,\alpha,\uparrow}n_{d,\alpha,\downarrow}, \nonumber
\end{eqnarray}
where $d^{\dagger}_{ \alpha,\sigma}$, $d^{}_{ \alpha,\sigma}$, are creation and
annihilation operators for an electron at the impurity site in channel
$\alpha$, where $\alpha=1,2$,  and spin
component
$\sigma=\uparrow,\downarrow$.  
The creation and annihilation operators $c^{\dagger}_{k,\alpha,\sigma}$, $c^{}_{k,\alpha,\sigma}$ are
for  partial wave conduction electrons with energy
$\epsilon_{k,\alpha}$ in channel $\alpha$, each with a bandwidth $2D$, with $D=1$. 
The Hamiltonian ${\cal H}_{12}$  we take to have an antiferromagnetic exchange term $J$
and a direct interaction $U_{12}$
between the two impurities,
\begin{equation}
{\cal H}_{12}= 
2J {\bf S}_{d,1}\cdot{\bf S}_{d,2}+U_{12}\sum_{\sigma,\sigma'}n_{d,1,\sigma}n_{d,2,\sigma'}.  
\label{model2c}
\end{equation}
For simplicity we consider identical dots  so we can drop the index $\alpha$ for the impurities.
\par 
The model has been well studied, in this\cite{NCH12a,NCH12b} and earlier forms where
the impurities are described by Kondo models\cite{JV87,JVW88,JV89,AL92,ALJ95,DF04,ZCSV06}.  The main focus of these studies
has been the quantum critical point which occurs at a critical coupling  $J=J_c$
on increasing $J$. For $J<J_c$ any magnetic screening of the impurities is via
the conduction electrons in their respective baths, but for  $J>J_c$,
the impurities are screened locally by the interaction between them.
Here we are concerned with the phases on either side of this transition
for the model away from particle-hole symmetry.\par
 
\begin{figure}[!htbp]
   \begin{center}
     \includegraphics[width=0.3\textwidth]{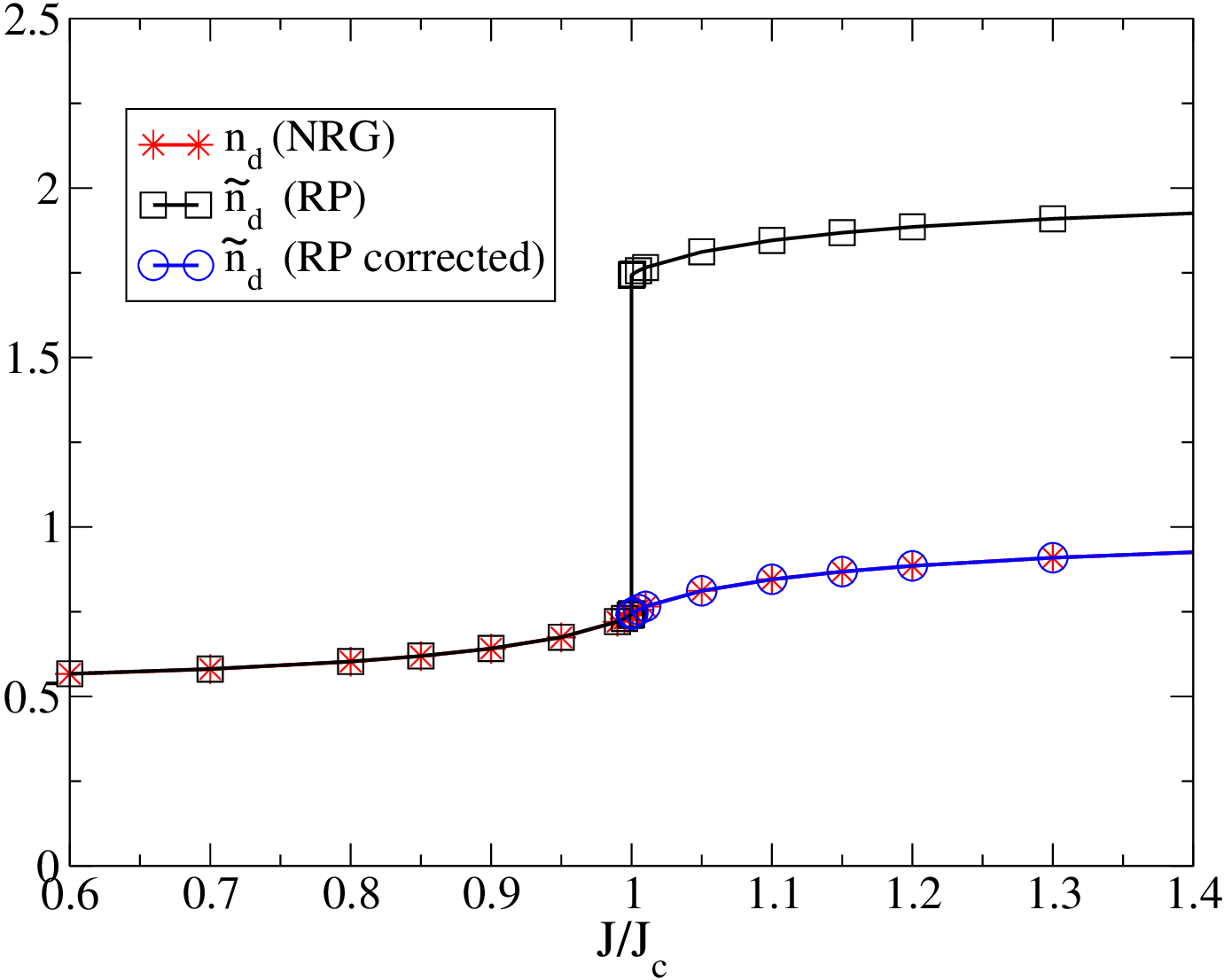}
     \caption{(Color online)  A plot of the impurity site occupation number
 $n_d$, as calculated directly from the NRG, as a function of $J/J_c$, compared with  $\tilde n_d$ (RP) from the Friedel
sum rule and as corrected with the Luttinger integral, for  $\epsilon_d/\pi\Delta=0.159$, $\pi\Delta=0.01$,  $U /\pi\Delta=0.5$ and $J_c=5.4401763\times 10^{-3}$.}
\label{compn1}
   \end{center}
 \end{figure}

The NRG low energy fixed point and the leading irrelevant terms of this model for 
a Fermi liquid fixed point  can be
analysed by replacing the parameters $\epsilon_d$, $V_{k}$,
$U$, $J$ and $U_{12}$,  by renormalized values,   $\tilde\epsilon_d$, $\tilde V_{k}$,
$\tilde U$, $\tilde J$ and $\tilde U_{12}$ with the additional proviso that all two-body interaction terms
have to be normal ordered. Though we take $U_{12}=0$ in all cases considered here there are finite values 
of $\tilde U_{12}$ to be taken into account in general.
The renormalized parameters (RP) can be deduced from the single
particle and two-particle excitations on the approach to the fixed point as has been described
elsewhere\cite{HOM04}.  The phase shift $\eta$ in terms of the free quasiparticles is given by
\begin{equation} \eta=\frac{\pi}{2}- {\rm tan}^{-1}\left(\frac{\tilde\epsilon_d}{\tilde\Delta}\right),\label{eta}\end{equation}
from which we deduce a value for $\tilde n_d$, the total quasiparticle occupation number per impurity site, from the relation $\tilde n_d=2\eta/\pi$.
The results are shown in Fig. \ref{compn1} as a function of $J/J_c$ for the particular parameter
set,  $\epsilon_d/\pi\Delta=0.159$, $\pi\Delta=0.01$,
and  $U /\pi\Delta=0.5$. This is compared with the total occupation value on each dot $n_d$ as calculated
directly from an NRG calculation  from the expectation value of $\sum_\sigma d^{\dagger}_{\sigma}d^{}_{\sigma}$
in the ground state. For $J<J_c$ there is a very precise agreement between the values of  $\tilde n_d$ and  $n_d$. At $J=J_c$ there is a sudden jump in the value of  $\tilde n_d$ by 1, which 
corresponds to a jump in the phase shift $\eta$ by $\pi/2$. This persists for $J>J_c$ such that the value of
 $\tilde n_d$ exceeds  $n_d$ by 1. The phase shift of $\pi/2$ cannot be accounted for by a jump to another
branch of the arctan; it suggests that the more general Luttinger-Friedel sum rule given in Eqn. (\ref{lfsr})
should be used in calculating $\tilde n_d$.

 \par
\begin{figure}[!htbp]
   \begin{center}
     \includegraphics[width=0.35\textwidth]{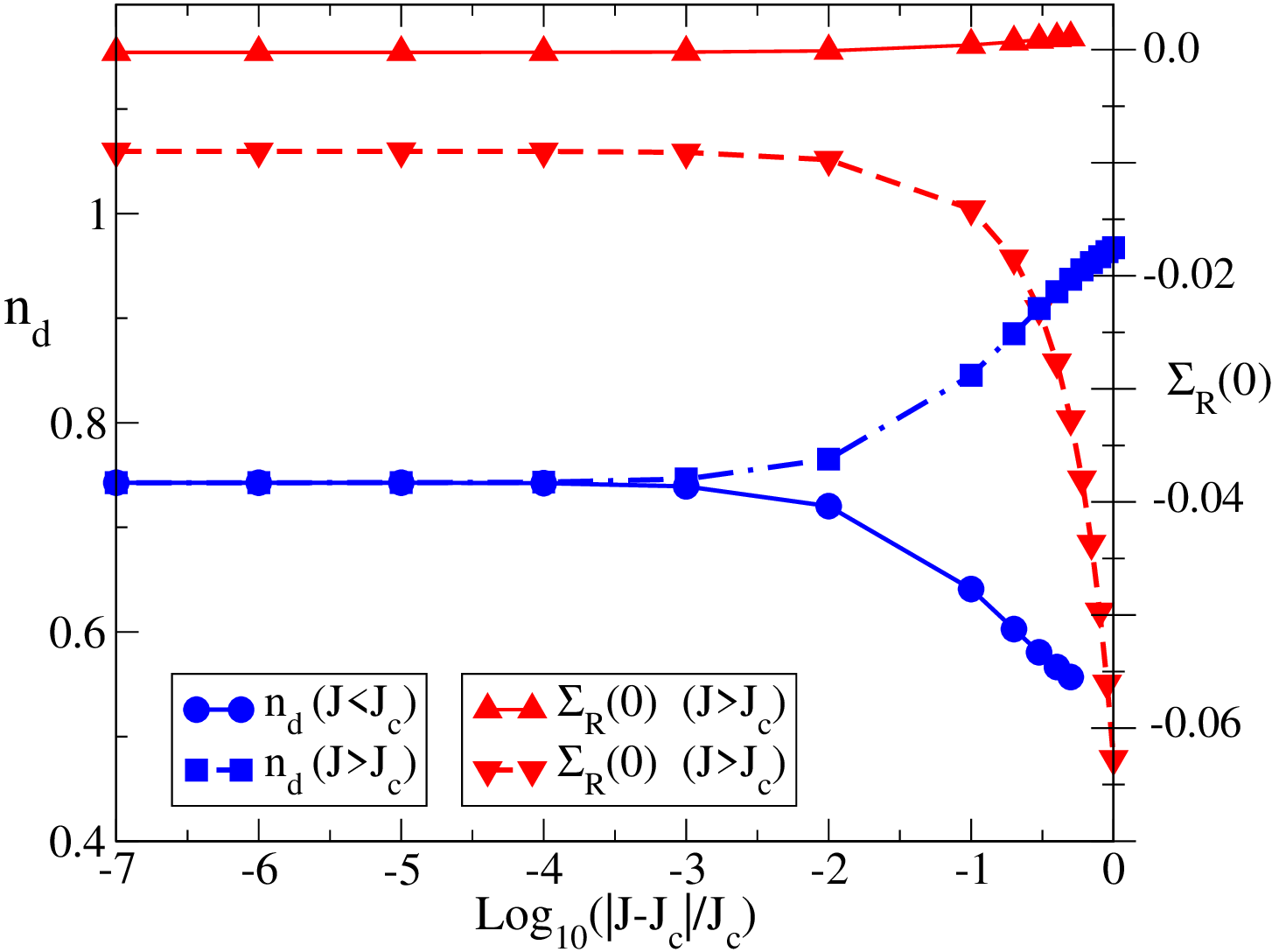}
     \caption{(Color online) A plot of the occupation number per site $n_d$ and the real part of the self-energy   $\Sigma_R(\omega)$ at $\omega=0$,     as a function of ${\rm Log_{10}}(|J-J_c|/J_c)$ for the parameter set in Fig. \ref{compn1}.
} 
     \label{sig_nd_plot}
   \end{center}
 \end{figure}

\begin{figure}[!htbp]
   \begin{center}
     \includegraphics[width=0.3\textwidth]{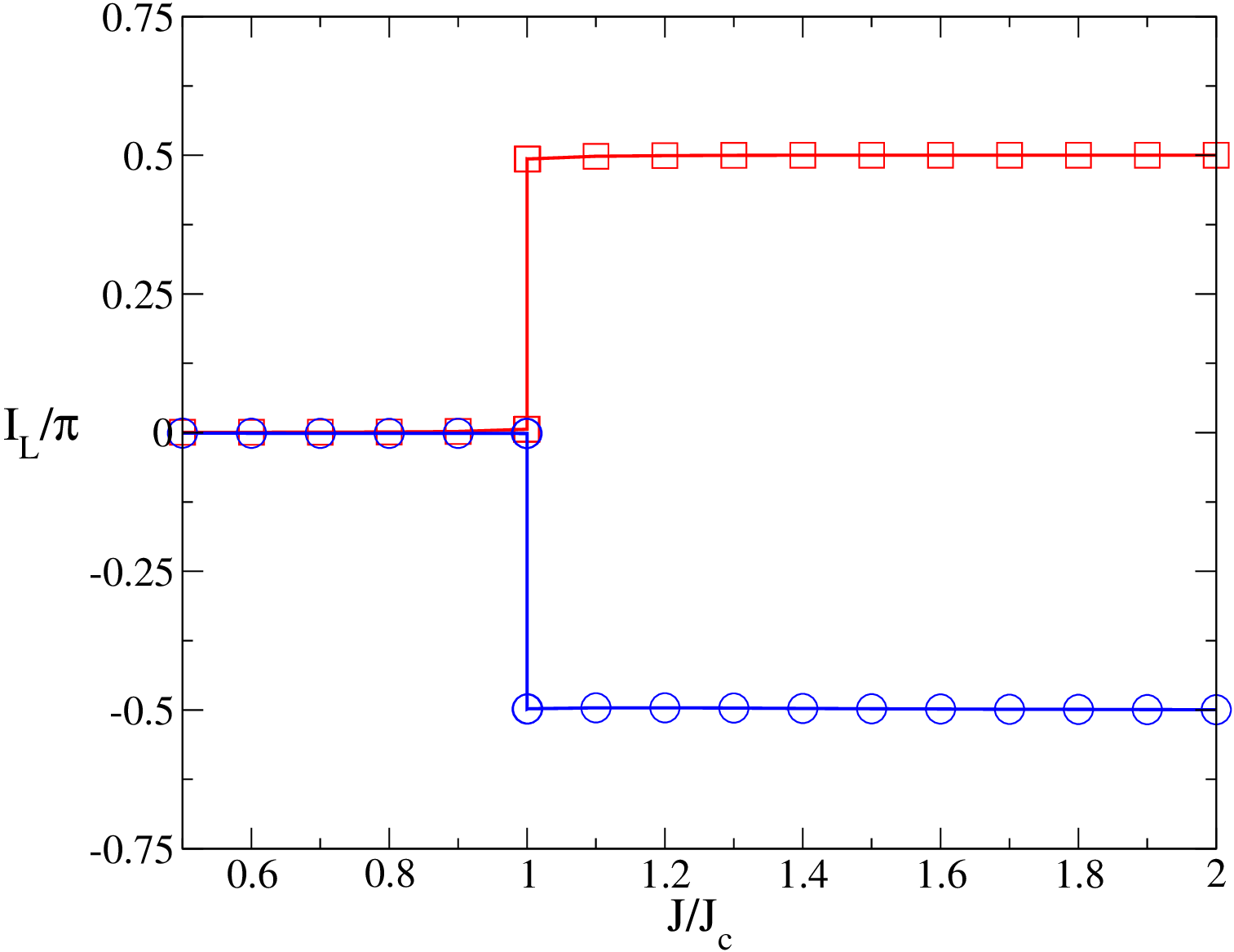}
     \caption{(Color online) A plot of  the Luttinger integral $I_{\rm L}/\pi$
    as a function of $J/J_c$ for the parameter set in Fig. \ref{compn1} (circles) and the set, $\epsilon_d/\pi\Delta=-1.0$, $\pi\Delta=0.01$,
  $U=0$ and $J_c=1.5126323\times 10^{-2}$ (squares).
} 
     \label{Lut}
   \end{center}
 \end{figure}

 
 To test this result we carry out an alternative direct calculation of $I_{\rm L}$ using the NRG results for  the self-energy and Green's function for one of the
impurity sites over the same range. We can rewrite the expression for $I_{\rm L}$ in the form,
\begin{equation}
I_{\rm L}=-\int_{-\infty}^0 {\rm Im}\,G_d(\omega)d\omega-\frac{2}{\pi}\left[\frac{\pi}{2}-{\rm tan}^{-1}\left(\frac{\epsilon_d+\Sigma_R(0)}{\Delta}\right)\right].\label{integral}\end{equation}
The results for $n_d$ and $\Sigma_R(0)$ across the transition are shown in Fig. \ref{sig_nd_plot} for the  parameter set used in Fig. \ref{compn1}. They  show clearly that the non-zero value of the Luttinger integral
$I_{\rm L}$ arises from the disconinuity in  $\Sigma_R(0)$ as the value of $n_d$ as calculated from the
integral term on the right hand side of Eqn. (\ref{integral}) is continuous across the transition.
The corresponding result for $I_{\rm L}$ is shown in Fig. \ref{Lut} showing that  $I_{\rm L}=\pi/2$ for all values with  $J>J_c$. Also shown are the results for a second parameter  
set,  $\epsilon_d/\pi\Delta=-1.0$, $\pi\Delta=0.01$, 
and  $U=0$ ($J_c=1.5126323\times 10^{-2}$),  where the impurity level lies below the Fermi level $\epsilon_d<0$,
giving 
$I_{\rm L}=-\pi/2$ for $J>J_c$.  When taking these values into account on applying the more general Luttinger-Friedel sum rule in Eqn. (\ref{lfsr}) the
relation   $\tilde n_d= n_d$ is restored.\par
 
  \begin{figure}[!htbp]
   \begin{center}
     \includegraphics[width=0.3\textwidth]{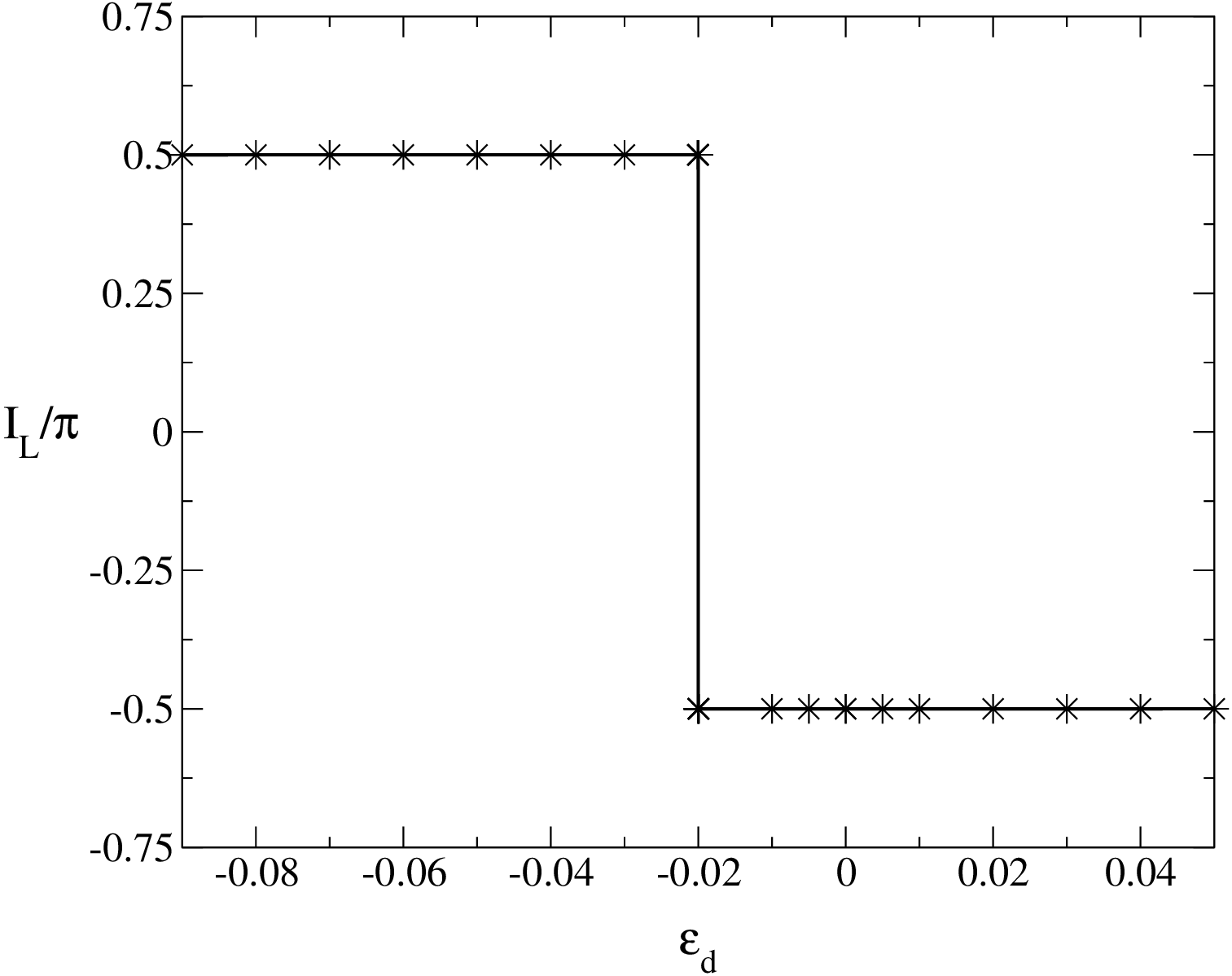}
     \caption{(Color online) A plot of  the Luttinger integral $I_{\rm L}/\pi$
    as a function of $\epsilon_d$ for the parameter set with $J/\pi\Delta=8$, $U/\pi\Delta=4$ and $\pi\Delta=0.01$.
} 
     \label{Epsilon}
   \end{center}
 \end{figure}
\par\bigskip

To test this behavior more generally we calculated  $I_{\rm L}/\pi$ for the parameter set $J/\pi\Delta=8$, $U/\pi\Delta=4$, $\pi\Delta=0.01$, and varied  $\epsilon_d$, where $J>J_c$ in all cases.
The results for $I_{\rm L}/\pi$ are shown in Fig. \ref{Epsilon} plotted as a function of $\epsilon_d$. In all cases $J>J_c$, we find a constant value $I_{\rm L}/\pi=1/2$ 
 over range  $\epsilon_d< -U/2$ and  $I_{\rm L}/\pi=-1/2$ 
 over range  $\epsilon_d> -U/2$, where the change of sign is at the point with particle-hole symmetry.
We conclude that $I_{\rm L}$ takes constant values in the different phases.

\par\bigskip
  \begin{figure}[!htbp]
   \begin{center}
     \includegraphics[width=0.32\textwidth]{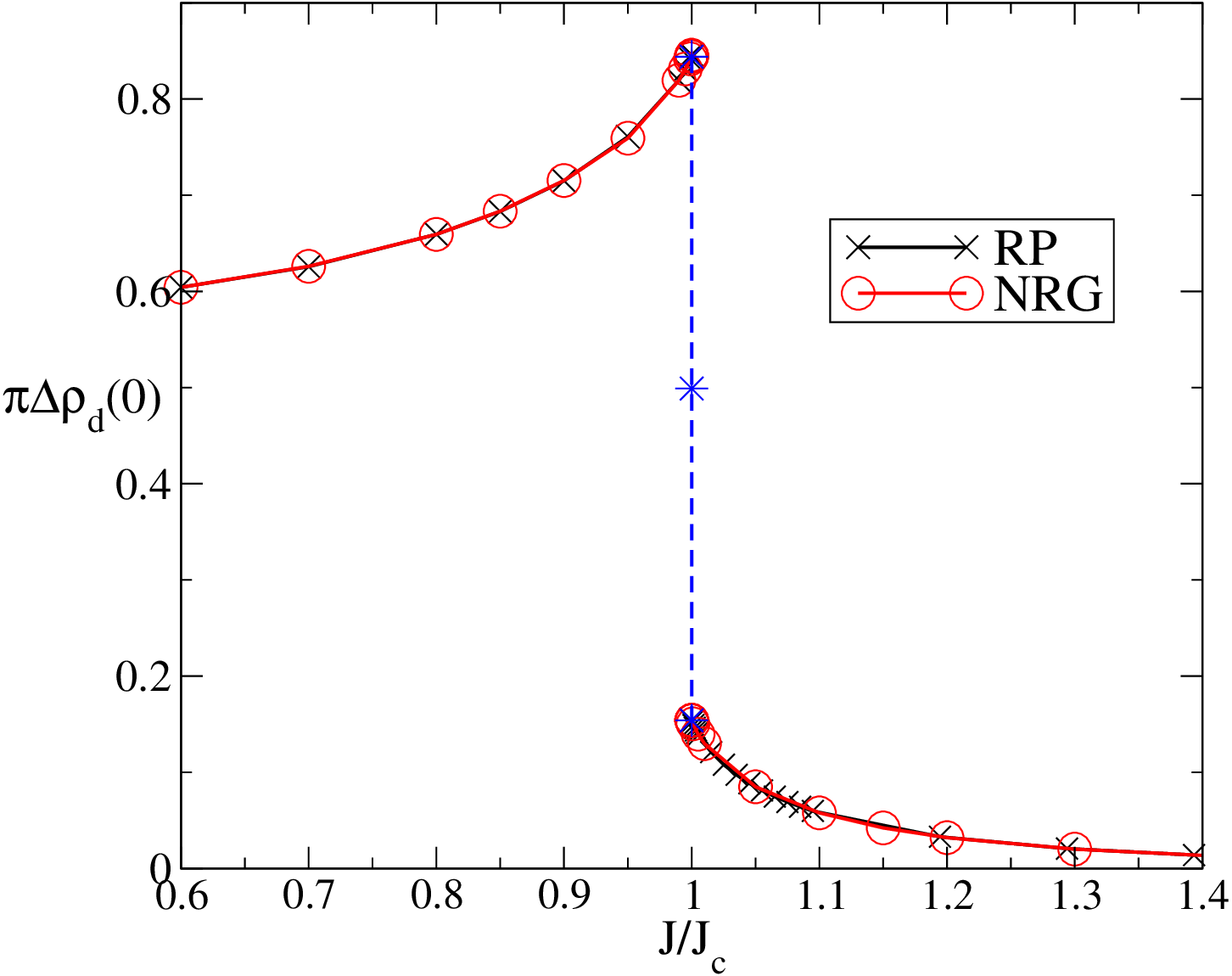}
     \caption{(Color online) A plot of $\pi\Delta\rho_d(0)$
    as a function of $J/J_c$ for the parameter set in Fig. \ref{compn1} as calculated from  the renormalized parameters (crosses) and from the NRG calculated spectral density  (circles).
} 
     \label{comprho}
   \end{center}
 \end{figure}
\bigskip
The jump in the  phase shift  of $\pi/2$, from $J_-=J_c-\delta$ to $J_+=J_c+\delta$, $\delta\to 0^{+}$, from Eqn. (\ref{eta}) implies a discontinuity in $\tilde\epsilon_d/\tilde\Delta$ such that
\begin{equation} \left(\frac{\tilde\epsilon_d}{\tilde\Delta}\right)_+ \left(\frac{\tilde\epsilon_d}{\tilde\Delta}\right)_-=-1,\label{disc}\end{equation}
or equivalently a discontinuity in the value of $\Sigma(0)$. In the Luttinger-Friedel sum rule this is compensated
by the jump in the Luttinger integral to $\pm\pi/2$, so that the value of $n_d$ is continuous through
the transition. The sudden discontinuity in $\Sigma(0)$ is however reflected in the spectral density of states $\rho_d(\omega)$  at the impurity
site at the Fermi level $\omega=0$. In terms of the phase shift $\rho_d(0)$ is given by
\begin{equation}\rho_d(0)=\frac{\sin^2(\eta)}{\pi\Delta}
=\frac{1}{\pi\Delta}\,
\frac{\tilde\Delta^2}{\tilde\epsilon_d^2+\tilde\Delta^2}.\label{rho0}\end{equation}
We can calculate this quantity from Eqn. (\ref{rho0}) using  renormalized parameters as deduced from the
low energy fixed point or directly from the self-energy $\Sigma(\omega)$ as calculated via the NRG.
In Fig. \ref{comprho} we give the results for $\pi\Delta\rho_d(0)$ as a function of $J/J_c$ for the parameter set in Fig. \ref{compn1}. We see complete agreement between the two sets of results, confirming the interpretation of the state in the regime $J>J_c$ as a Fermi liquid. The mid-point of the discontinuity, indicated by a star in Fig \ref{comprho}, corresponds to $\rho_d(0)=1/2\pi\Delta$, and seems to be a general feature independent of the particular parameter set chosen.  
\par
\begin{figure}[!htbp]
   \begin{center}
     \includegraphics[width=0.35\textwidth]{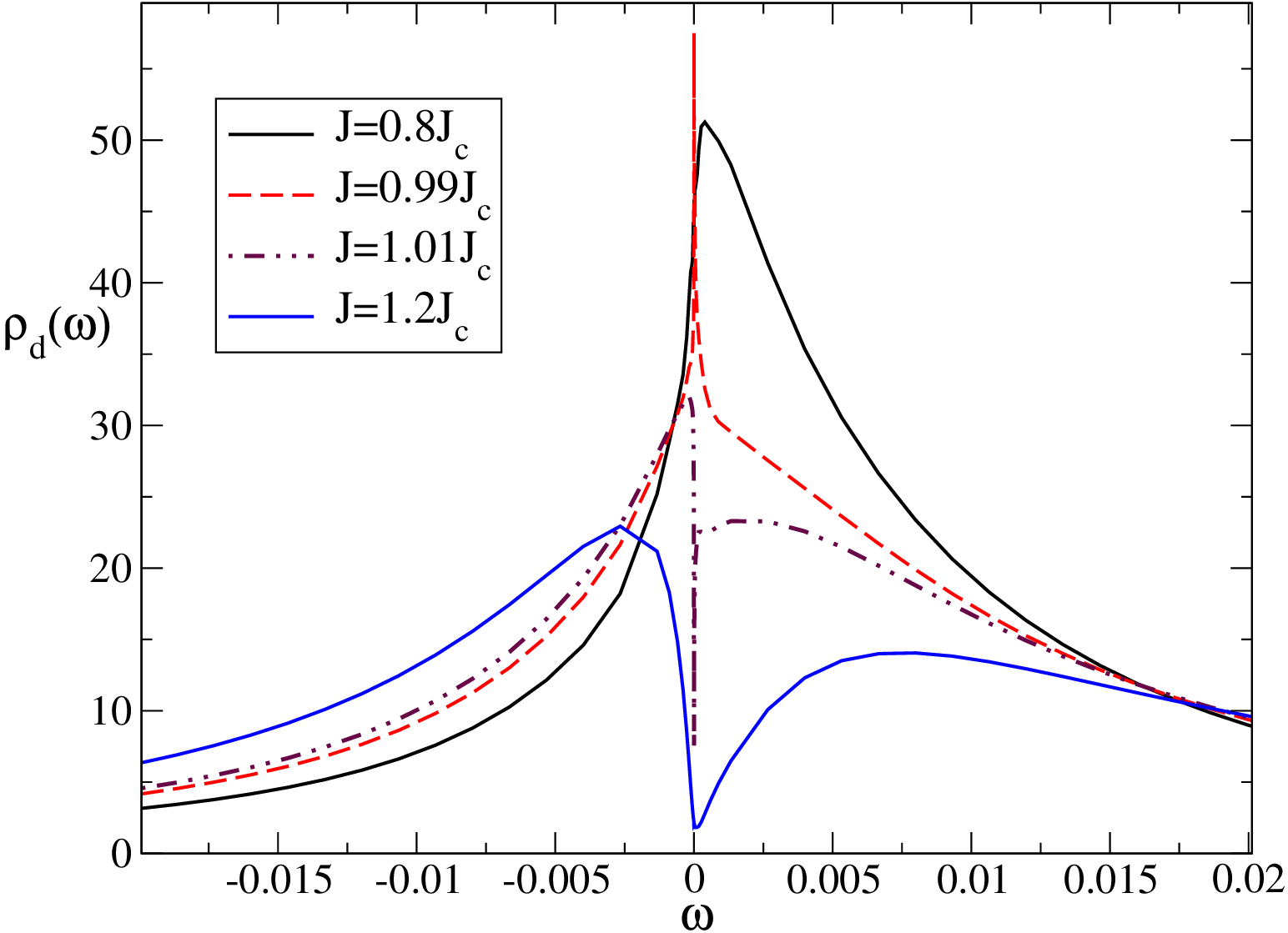}
     \caption{(Color online) A plot of $\rho_d(\omega)$ 
    as a function of $\omega$ for the parameter set in Fig. \ref{compn1} with values of $J/J_c=0.8,0.99,1.01,1.2$.
} 
     \label{rhos}
   \end{center}
 \end{figure}

\bigskip
Apart from the sudden jump in the value of $\rho_d(0)$ at $J=J_c$, there is a continuous redistribution of
the spectral weight   $\rho_d(\omega)$ as $J$ varies through the transition region. In Fig. \ref{rhos}
we show this change for the parameter set in Fig. \ref{compn1} by comparing the forms for   $\rho_d(\omega)$
for $J/J_c=0.8,0.99,1.01,1.2$. For $J=0.8J_c$ there is a single broad peak above the Fermi level, which  becomes very narrow and shifts to just above the Fermi level at $J=0.99J_c$. After the transition for  $J=1.01J_c$
there is a sudden drop in the spectral density at the Fermi level and a peak just below the Fermi level.
For  $J=1.2J_c$ the peak has shifted to lower energies and broadened with a distinct local minimum in $\rho_d(\omega)$ at the Fermi level. The form of the spectral density in the immediate region of the Fermi level is to a good approximation given by the spectral density due to the free quasiparticles, $\tilde\rho_d(\omega)=\tilde\Delta/\pi((\omega-\tilde\epsilon_d)^2+\tilde\Delta^2)$, when multiplied by the quasiparticle weight factor $z=\tilde\Delta/\Delta$, reflecting the Fermi liquid nature of the low lying excitations.  As $J\to J_c$, $\tilde\epsilon_d\to 0$ and $\tilde\Delta\to 0$, this quasiparticle expression gives the narrowing of the peak on the approach to the transition. The discontinuity in $\tilde\epsilon_d$ at $J=J_c$
and  change of sign from Eqn. (\ref{disc}) gives the shift of the peak across the Fermi level. 
\par

\par\bigskip

\begin{figure}[!htbp]
   \begin{center}
     \includegraphics[width=0.28\textwidth]{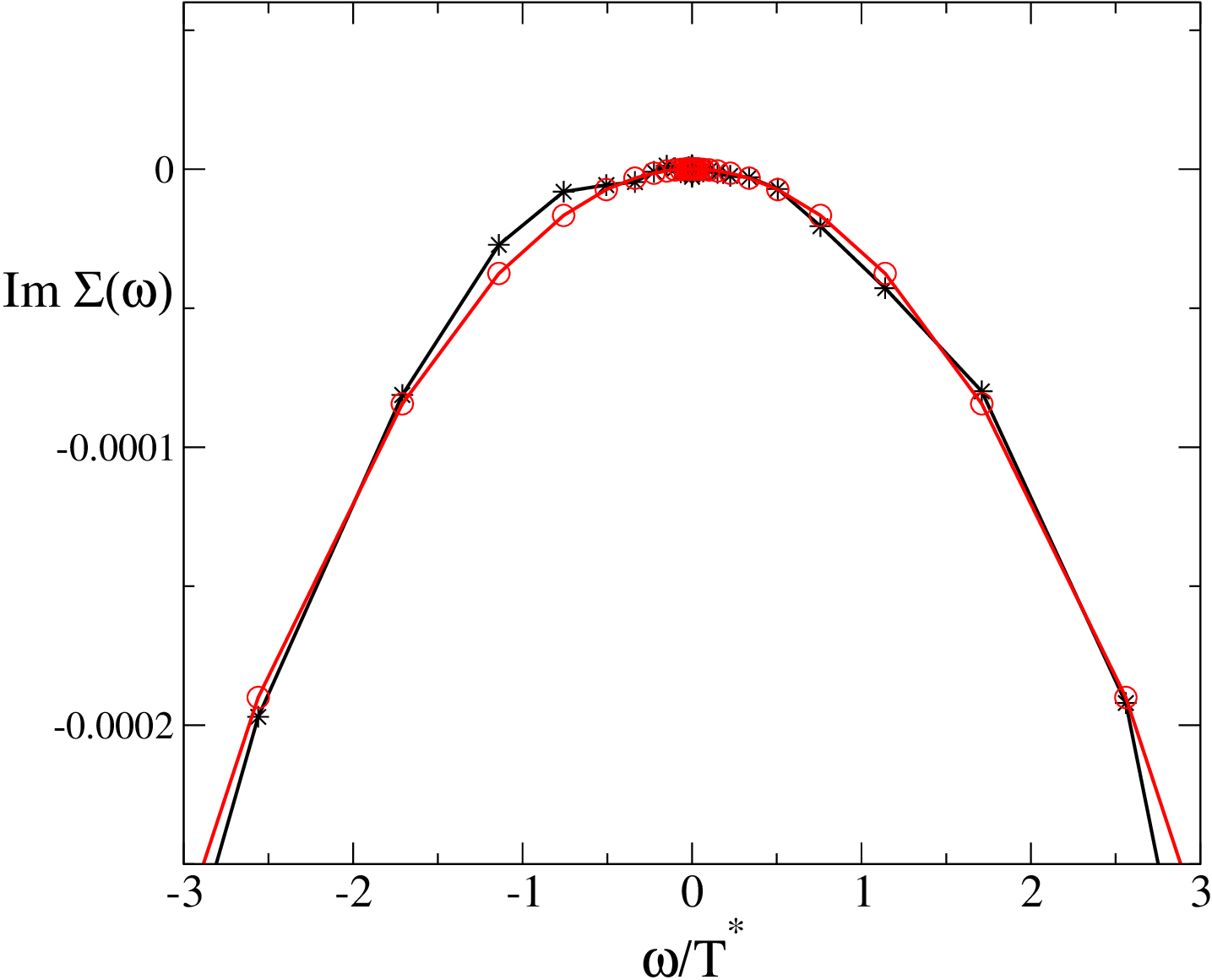}
     \caption{(Color online) A plot of ${\rm Im}\Sigma(\omega)$
    as a function of $\omega/T^*$ for the parameter set in Fig. \ref{compn1}  from  the NRG results (stars) with a quadratic fit (circles)  for $J=2J_c$ and $T^*=\pi\tilde\Delta/4=9.14553\times 10^{-5}$.
} 
     \label{rhofit}
   \end{center}
 \end{figure}

Finally in Fig. \ref{rhofit} we give the imaginary part of the self-energy $\Sigma(\omega)$  as a function of
$\omega/T^*$, where $T^*$  is the renormalized energy scale $T^*=\pi\tilde\Delta/4$. Here for Fermi liquid behavior, as in the single impurity Anderson model, we expect an $\omega^2$ form on the scale $\omega<T^*$. There are some inaccuracies in calculating this quantity from an NRG calculation due to broadening of discrete data, but there is a very reasonable fit to the quadratic form as given in the plot. 
\par

We have established that in this model, away from particle-hole symmetry, we have three Fermi liquid phases.
Only one of them has the expected value $I_{\rm L}=0$  for the Luttinger integral. The other two have
constant values of  $I_{\rm L}$ with  $I_{\rm L}=\pi/2$ or  $I_{\rm L}=-\pi/2$. As the case with $I_{\rm L}=0$
includes the case $J=0$ and the single impurity Anderson model, it fits the condition in some definitions
of a Fermi liquid that the states of the interacting system correspond to an adiabatic evolution from those
of the non-interacting system. This is not the case for the phases with  $I_{\rm L}=\pm\pi/2$, but nevertheless
they satisfy all the other usual requirements of a Fermi liquid; well defined low energy quasiparticles,
with non-singular scattering leading to the usual $\omega^2$ terms, and consequent $T^2$ low temperature
behavior.
The case with particle-hole symmetry is different. Though there is a sudden change of phase shift by $\frac{\pi}{2}$ at $J=J_c$, for $J>J_c$ 
 we find the self-energy has a simple pole, $\Sigma(\omega)\sim\frac{1}{\omega}$ as $\omega\to 0$, and consequently  the spectral density goes to zero at the Fermi level.
 \par
 The Wilson ratios for a spin, charge, staggered spin and charge, in the Fermi liquid regimes on both sides of the transition at $J=J_c$ were calculated in earlier work\cite{NCH12a,NCH12b}
from the  renormalized parameters for the quasiparticles, and were in complete agreement with exact results found in essentially the same model studied by De Leo and Fabrizio\cite{DF04}.\par

 The different  Fermi liquid phases can be classified  by the quantum number ${2 I_{\rm L}}/{\pi}$, which is not associated with any symmetry. This could give a general explanation as to puzzling question as to why the transition
 in this model is so robust, existing not only away from particle-hole symmetry but also for $U=0$.
As this quantum number cannot change continuously at any transition between these phases, it implies that
 the transition at $J=J_c$ is essentially a topological one. Our results also raise the question as to whether
the Luttinger integral can take similar values and modify the standard Luttinger relation in strong correlation lattice models, such as the t-J model. 
 \par
 \par\bigskip
One of us (Y.N.)  acknowledges the support by JSPS KAKENHI Grant No.15K05181.
\bibliography{artikel,biblio1}

\end{document}